\documentclass[preprint,preprintnumbers,amsmath,amssymb]{revtex4}

\usepackage{epsfig}
\usepackage{tikz}
\usepackage{graphicx}
\usepackage{dcolumn}
\usepackage{bm}
\usepackage{url}
\usepackage{xcolor}

\begin{document}



\title{Disentangling high order effects in the transfer entropy}

\author{Sebastiano Stramaglia$^{1}$, Luca Faes$^{2}$, Jesus M Cortes$^{3}$, and Daniele Marinazzo$^{4}$}

\affiliation{$^1$ Dipartimento Interateneo di Fisica, Universit\`a degli Studi di Bari Aldo Moro, and INFN, Sezione di Bari, 70126 Bari, Italy\\}

\affiliation{$^2$ Dipartimento di Ingegneria, Universit\`a di Palermo, 90128, Palermo, Italy \\}

\affiliation{$^3$ Biocruces-Bizkaia Health Research Institute, Barakaldo, Spain\\
Biomedical Research Doctorate Program, University of the Basque Country, Leioa, Spain\\
Department of Cell Biology and Histology, University of the Basque Country, Leioa, Spain\\
IKERBASQUE Basque Foundation for Science, Bilbao, Spain \\}

\affiliation{$^4$  Department of Data Analysis, Ghent University, Ghent, Belgium\\}

\date{\today}

\begin{abstract}

Transfer Entropy (TE), the primary method for determining directed information flow within a network system, can exhibit bias—either in deficiency or excess—during both pairwise and conditioned calculations, owing to high-order dependencies among the dynamic processes under consideration and the remaining processes in the system used for conditioning. Here, we propose a novel approach. Instead of conditioning TE on all network processes except the driver and target, as in its fully conditioned version, or not conditioning at all, as in the pairwise approach, our method searches for both the multiplets of variables that maximize information flow and those that minimize it. This provides a decomposition of TE into unique, redundant, and synergistic atoms. Our approach enables the quantification of the relative importance of high-order effects compared to pure two-body effects in information transfer between two processes, while also highlighting the processes that contribute to building these high-order effects alongside the driver. We demonstrate the application of our approach in climatology by analyzing data from El Ni\~{n}o and the Southern Oscillation.

\end{abstract}\maketitle

\maketitle
A central task in analyzing complex systems is to understand the joint dynamics of its components. Granger causality (GC) \cite{granger} and transfer entropy (TE) \cite{schreiber,book_te} are widely used tools to detect and quantify statistical relationships between random processes mapping the evolution of coupled dynamic systems over time in terms of reduction of variance and surprise, respectively. For Gaussian systems, GC and TE are equivalent \cite{barnett2009}; a symbolic version of it has been developed in \cite{lehnertz}. 

The target properties of these statistical dependencies have been called "information flow" or "causality", names which will be used in this paper to indicate a measured effect \cite{barbar}; this clarification pairs an important distinction that needs indeed to be made between mechanisms and behaviors in this context \cite{rosas}.

In recent years, alongside the growing interest in high-order interactions \cite{battiston,rosas}, there has been increasing attention devoted to the emergent properties of complex systems, which manifest through high-order behaviors sought in observed data, moving beyond traditional dyadic descriptions. A key framework in this literature is the partial information decomposition (PID) \cite{pid} and its subsequent developments \cite{lizier}, which utilize information-theoretic tools to reveal high-order dependencies among groups of three or more random variables and describe their synergistic or redundant nature. Within this framework, redundancy refers to information retrievable from multiple sources, while synergy refers to statistical relationships existing within the whole system that cannot be observed in its individual parts.
Importantly, while the PID was originally proposed for sets of random variables, it was then generalized to random processes \cite{faes2017multiscale}.

It is worth mentioning that in \cite{cruc}, a critique of TE has been raised: if one believes that a dyadic network accurately models a complex system, then one implicitly assumes that polyadic relationships are either unimportant or nonexistent. However, assessing the relative importance of these high-order interactions compared to dyadic relations remains an open problem. Moreover, in \cite{stramaglia}, the influence of synergy and redundancy on the inference of information flow between two subsystems of a complex network, in terms of GC, has been studied. This research demonstrates that both pairwise and fully conditioned GC analyses encounter challenges in the presence of synergy or redundancy in time series data.

The question we address here is: how can we compare the pure dyadic influence with the many-body effects due to the remaining processes in a network, given a driving random process and a target process? Within the framework of partial conditioning in multivariate data sets \cite{partial}, we propose a straightforward decomposition of the maximal information flow from the driver to the target into three positive components: unique, redundant, and synergistic TEs. The magnitudes of these components quantify the relative importance of high-order effects and pure dyadic effects in the influence from the driver to the target. Moreover, the proposed analysis highlights the processes that, along with the driver, contribute to synergistic (or redundant) effects on the target.

We start observing the following fundamental fact behind any approach to PID:
for any three (scalar or vector) random variables $a$, $b$ and $c$, being $H$ the entropy and $I$ the mutual information, we will always have $H(a)\ge H(a|c)$ but we do not necessarily have  $I(a;b)\ge I(a;b|c)$.
In fact, conditioning on $c$ can either reduce the information shared between $a$ and $b$ or increase it: the two cases are related respectively to the prevalence of redundancy ($I(a;b) > I(a;b|c)$) or to the prevalence of synergy ($I(a;b) < I(a;b|c)$) among the three variables.
Now, suppose that $c$ is a subset of all the variables at hand which describe the {\it environment} around the pair $(a,b)$; it is intuitive that searching for $c_{min}$ minimizing $I(a;b|c)$ provides the amount of redundancy that the {\it environment} shares with the pair, i.e.  $I(a;b)-I(a;b|c_{min})$. On the other hand, searching for $c_{max}$ maximizing $I(a;b|c)$ leads to the amount of synergy that the {\it environment} provides in terms of the increase of mutual information $I(a;b|c_{max})-I(a;b)$. 
In what follows we will apply this heuristics to the transfer entropy. 

Given a zero-mean stationary two-dimensional Markov process consisting of the scalar processes $X$ and $Y$, we aim at evaluating the transfer entropy $X\to Y$. Denoting the present state of the target as $Y_t$, the vector of the target's past variables as $Y_{<t} = [Y_{t-1}\ldots Y_{t-p}]^T$, and the vector of the driver's past variables as $X_{<t} = [X_{t-1}\ldots X_{t-p}]^T$ , $p$ being the order of the process,
the pairwise TE, $T_{p}$, is defined as:
\begin{equation}
T_{p}=T_{X\rightarrow Y}= I\left(Y_t;X_{<t}|Y_{<t}\right),\\
\label{eq:pairwise}
\end{equation}
where $I(\cdot;\cdot|\cdot)$ indicates the conditional mutual information.
Suppose now that we also simultaneously measure $n$ other processes $\boldsymbol{Z}=\{Z_1,Z_2,\ldots,Z_n\}$.
The fully conditioned TE, $T_{f}$, is defined as:
\begin{equation}
T_{f}=T_{X\rightarrow Y|\boldsymbol{Z}}= I\left(Y_t;X_{<t}|Y_{<t},Z_{1,<t},\ldots,Z_{n,<t}\right)
\label{eq:multivariate}
\end{equation}
Now, instead of fully conditioning on all processes, we condition only on a subset in $\boldsymbol{Z}$, i.e. we compute $T_{\alpha}=T_{X\rightarrow Y|\boldsymbol{Z}_{\alpha}}$, where $\boldsymbol{Z}_{\alpha}$,
$\alpha \subset \{1,\ldots,n\}$, is an element of the powerset of $\{Z_1,Z_2,\ldots,Z_n\}$.
We denote as ${\boldsymbol{Z}_m}$ the subset of processes in $\boldsymbol{Z}$ which minimizes $T_{\alpha}$, and the corresponding value of the TE will be denoted as $T_m$:
\begin{equation}
T_{m}=T_{X\rightarrow Y|\boldsymbol{Z}_m}= I\left(Y_t;X_{<t}|Y_{<t},\boldsymbol{Z}_{m,<t}\right)
\label{eq:minimum}
\end{equation}

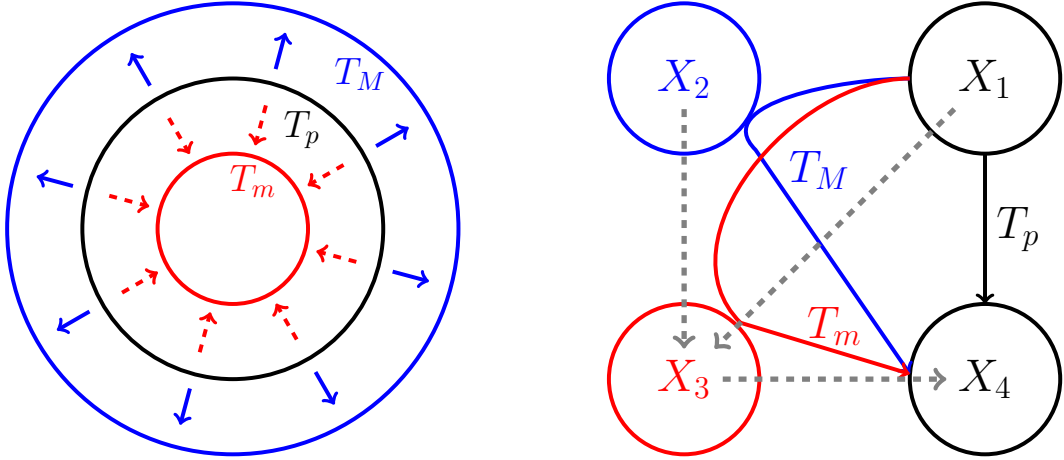
\begin{figure}[ht!]
\centering
\begin{tikzpicture}[scale=1.0]

\def\innerRadius{1cm}
\def\middleRadius{2cm}
\def\outerRadius{3cm}
\def\smallRadius{1cm}

\draw[red, line width=1.5pt] (0,0) circle (\innerRadius) node[above, font=\large] at (45:.38*\innerRadius) {$T_m$};
\draw[line width=1.5pt] (0,0) circle (\middleRadius) node[above, font=\large] at (45:.65*\middleRadius) {$T_p$};
\draw[blue, line width=1.5pt] (0,0) circle (\outerRadius) node[above, font=\large] at (45:.8*\outerRadius) {$T_M$};

\foreach \angle in {30,75,...,360}
  \draw[red, line width=1.5pt, dashed, ->] ({.85*\middleRadius*cos(\angle)}, {.85*\middleRadius*sin(\angle)}) -- ({1.15*\innerRadius*cos(\angle)}, {1.15*\innerRadius*sin(\angle)});

\foreach \angle in {30,75,...,360}
  \draw[blue, line width=1.5pt, ->] ({1.1*\middleRadius*cos(\angle)}, {1.1*\middleRadius*sin(\angle)}) -- ({0.9*\outerRadius*cos(\angle)}, {0.9*\outerRadius*sin(\angle)});

\begin{scope}[shift={(8,0)}]

\draw[blue, line width=1.5pt] (-2,2) circle (\smallRadius) node[ font=\Large] (2) {$X_2$};
\draw[red, line width=1.5pt] (-2,-2) circle (\smallRadius) node [ font=\Large] (3) {$X_3$};
\draw[black, line width=1.5pt] (2,2) circle (\smallRadius) node [ font=\Large](1) {$X_1$};
\draw[black, line width=1.5pt] (2,-2) circle (\smallRadius) node [ font=\Large](4) {$X_4$};

\draw[black, line width=1.5pt, ->] (2,1) -- (2,-1) node[midway, right, font=\Large] {$T_p$};
\draw[blue, line width=1.5pt, ->] (\smallRadius,2) to[out=180,in=135] (-1.05,1.05) -- (2-1,2-2*\smallRadius);
\draw[red, line width=1.5pt, ->] (\smallRadius,2) to[out=180,in=135] (-1.25,-1.25)  -- (2-1,2-2*\smallRadius);


\path [->,draw, dashed, color=gray, line width=2pt](1) edge node[left] {} (3);
\path [->,draw, dashed, color=gray, line width=2pt](2) edge node[left] {} (3);
\path [->,draw, dashed, color=gray, line width=2pt](3) edge node[left] {} (4);

\node[below, font=\Large, red] at (0,-.9) {$T_m$};

\node[above, font=\Large, blue] at (-.2,0.4) {$T_M$};

\end{scope}
\end{tikzpicture}
\caption{\textit{Left}: Representation of the pairwise Transfer Entropy $T_{p}$ and of the values obtained maximizing and minimizing $T_{\alpha}$, called $T_{M}$ and $T_{m}$ respectively. Redundant influences are depicted by red arrows, and synergistic influences by blue arrows. \\ \textit{Right}: Pairwise, synergistic, and redundant conditioning path are represented on a simple model with four nodes, with simulated couplings indicated by the dashed gray arrows.}
\label{fig:circles}
\end{figure}

Similarly, ${\boldsymbol{Z}_M}$ represents the subset of processes in $\boldsymbol{Z}$ that maximizes $T_{\alpha}$, with the corresponding value of the Transfer Entropy denoted as $T_M$:

\begin{equation}
T_{M}=T_{X\rightarrow Y|\boldsymbol{Z}_M}= I\left(Y_t;X_{<t}|Y_{<t},\boldsymbol{Z}_{M,<t}\right).
\label{eq:maximum}
\end{equation}
This decomposition is illustrated in the left panel of figure \ref{fig:circles}.

Given that the reduction of TE is associated with redundancy (R), we express this relationship as $R = T_p - T_m$, where the unique information (U) flowing from $X$ to $Y$ is identified with $T_m$.
On the other hand, synergistic information flow (S) signifies the increase in transfer entropy when additional variables are included in the set of conditioning variables. This relationship can be expressed as $S = T_M - T_p$.
It follows that:
\begin{equation}
T_{M}= S+R+U.\\
\label{eq:decomposition}
\end{equation}
In other words, the maximal information flow from $X$ to $Y$ can be decomposed into the sum of a unique contribution (U), representing a pure two-body effect, and synergistic and redundant contributions that describe higher-order modulations of the interdependence between $X$ and $Y$.

Conducting an exhaustive search for subsets ${\boldsymbol{Z}_m}$ and ${\boldsymbol{Z}_M}$ becomes unfeasible for large $n$. Therefore, we employ a greedy search strategy, wherein firstly we perform a search over all the processes for the first process to be tentatively used as a conditioner. Subsequently, one process is added at a time, to the previously selected ones, to construct the set of conditioning processes that either maximize or minimize the TE.
To establish a criterion for terminating the greedy search for conditioning processes that minimize (maximize) the TE, one can estimate the probability that the increase in $T_{\alpha}$ is lower (higher) than that corresponding to the inclusion of a randomized time series (obtained, e.g., by   iterative amplitude-adjusted Fourier Transform (IAAFT) \cite{iaaft} surrogates of the selected $Z$ time series). The selected process is thus added to the multiplet of conditioning processes when such probability is lower than a given threshold, after correction for multiple comparisons.

As a toy model (which in the distinction between mechanisms and behavior would constitute a model for the former, but which should be reasonably echoed by the latter), consider a system of four random processes, with $X_1$ and $X_2$ influencing $X_3$, whilst $X_3$ influences $X_4$, as depicted by the gray dashed arrows in the right panel of figure \ref{fig:circles}:
    \begin{equation}
    \begin{cases}
    X_{1,t}=\epsilon_{1,t}\\
    X_{2,t}=\epsilon_{2,t}\\   
    X_{3,t}=0.5\left(X_{1,t-1}+ X_{2,t-1}\right) + 0.1\;\epsilon_{3,t}\\    
   X_{4,t} = 0.9\;X_{3,t-1} + 0.1\;\epsilon_{4,t}
    \end{cases}
    \label{toy}
\end{equation}
where $\epsilon$'s are unit variance zero mean i.i.d. Gaussian processes. Taking $X_1$ as the driver and $X_4$ as the target, we easily obtain $T_p=0.32$, $T_m=0$ (with ${\boldsymbol{Z}_m}$ coinciding with $X_3$) and $T_M=1.25$ (with ${\boldsymbol{Z}_M}$ coinciding with $X_2$); it follows that $U=0$ (there is no pure two-body influence), while $R=0.32$ and $S=0.93$ are the redundant and synergistic TEs respectively. 
Noting that $X_1$, $X_2$ are colliders for $X_3$, and that the chain $X_1\to X_3\to X_4$ is a redundant circuit, it is easy to realize that these results are what one should expect on this example. We remark the output, on this toy model, for the fully conditioned TE is $T_f=0$ whilst for the pairwise TE it is $T_p=0.32$: both approaches fail to highlight  many body effects.

As an example of an application to a real dataset, we consider a case study in climate science, i.e. the influence of NINO34 (the East Central Tropical Pacific sea surface temperature anomaly, also called El Ni\~{n}o) on SOI 
(southern Oscillation Index, the standardized difference in surface air pressure between Tahiti and Darwin). These two indexes are crucial for the description of  El Ni\~{n}o and the Southern Oscillation (ENSO), a periodic fluctuation in sea surface temperature and the air pressure of the overlying atmosphere across the equatorial Pacific Ocean. ENSO is considered the most prominent interannual climate variability on Earth \cite{mcphaden}. Since the exact initiating causes of an ENSO warm or cool event are not fully understood, it is important to analyze the statistical relation between the two components of ENSO—atmospheric pressure (SOI) and SST (NINO34). 
The other climatic indexes that we consider here are AIR (All Indian Rainfall), AMO (Atlantic Multidecadal Oscillation), GMT (Global Mean Temperature anomaly), HURR (total number of hurricanes or named tropical storms in a given month in the Atlantic region), NOA (North Atlantic Oscillation of pressure anomalies over the Atlantic), NP (North Pacific pattern of sea level pressure), NTA (North Tropical Atlantic), PDO (Pacific Decadal Oscillation), QBO (Quasi-Biennial Oscillation), Sahel (Sahel Standardized Rainfall), and TSA (Tropical Southern Atlantic Index). All time series have been detrended and deseasonalized; globally, these thirteen monthly sampled time series coincide with those analyzed using an approach based on a linear approximation of the pairwise transfer entropy in \cite{clima}. 
It is worth stressing that these variables are not guaranteed to measure separate processes, and that latent factors are likely to be present. This further motivates an analysis that takes shared information into account.

We consider the period 1950–2016 (i.e., 792 data points) for which all the values of the thirteen time series are available, and adopt the assumption of gaussianity so as to identify transfer entropy with GC. We find that the pairwise TE, NINO34$\to$SOI, is $T_p =0.078$, using $p=2$ as the order the model fixed by minimum description length criterion. Then we find $T_m =0.074$ (with ${\bf Z_m}$ equal to PDO, see fig.1-top) and $T_M =0.094$ (with ${\bf Z_M}$ equal to the pair NTA and TSA, see fig.1-bottom): therefore in this case we have $S=0.016$, $R=0.004$ and $U=0.0.74$. Since $\frac{(R+S)}{T_M}= 0.2$, we conclude that about $20\%$ of the total information flow from NINO34 to SOI can be ascribed to many-body effects. Moreover, we note that in \cite{clima}, using a pairwise approach, eight  drivers of SOI were identified among the thirteen time series, including NINO34, NTA e TSA; moreover PDO was found to be target for both NINO34 and SOI. The proposed approach allows to identify which drivers play a role to produce high order effects on SOI in cooperation with NINO34. 

\begin{figure}[!ht] \centering
\includegraphics[width=.8\textwidth]{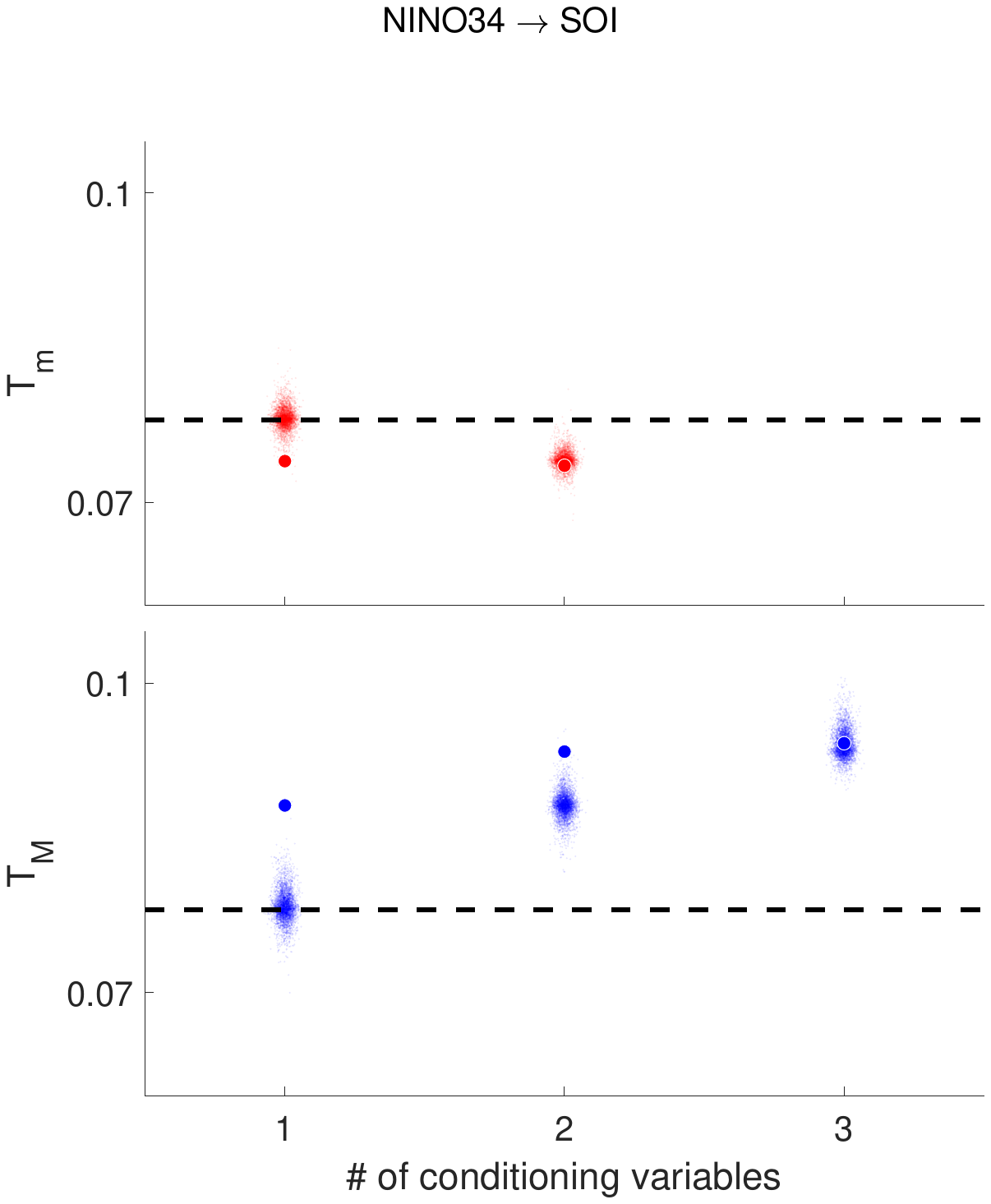}
\caption{Top: The transfer entropy NINO34$\to$SOI is depicted as a function of the number of conditioning variables along the redundant greedy search ($T_{m}$), together with the values obtained on 4000 surrogates of the added time series. Conditioning to a second variable results in values compatible with the null hypothesis, hence the set $\bf{Z_m}$ reduces to the first variables (PDO). The value of the pairwise Transfer Entropy ($T_{p}$) is indicated by the dashed black line. Bottom: As in top, but referring to the synergistic greedy search, resulting in $T_{M}$. Conditioning to a third variable results in values compatible with the null hypothesis, hence the set $\bf{Z_M}$ reduces to the first and second variables (NTA and TSA).}
\label{fig:atm}
\end{figure}

In summary, we have introduced a decomposition of Transfer Entropy (TE) that separates pure dyadic dependence from many-body effects resulting from interactions with other variables. The computational demand is not intensive, especially when employing a greedy search strategy to identify the optimal multiplet of conditioning variables. Applying this methodology to the interplay between NINO34 and SOI, two components of ENSO, revealed the presence of non-negligible redundant and synergistic high-order effects. We posit that this approach serves as a bridge between dyadic and polyadic methodologies, offering a complementary perspective to those focusing on the assessment of high-order effects \cite{pid,oinfo,doinfo}.

\clearpage
\section*{Data and code availability}
The code to simulate and analyze data is available at \url{https://github.com/danielemarinazzo/TE_decomposition}.
Climate data as described in \cite{clima} can be downloaded at the NOAA website https://psl.noaa.gov/data/climateindices/list/), with the exception of AIR, which is available via the Indian Institute of Tropical Meteorology (https://www.tropmet.res.in/).
\begin{acknowledgments}
LF and SS were supported by the project “HONEST - High-Order Dynamical Networks in Computational Neuroscience and Physiology: an Information-Theoretic Framework”, Italian Ministry of University and Research (funded by MUR, PRIN 2022, code 2022YMHNPY, CUP: B53D23003020006).
We thank Yves Rosseel and Niels Van Santen (Ghent University) for useful comments.
\end{acknowledgments}

\end{document}